\documentstyle[osa,manuscript]{revtex}

\begin{document}

\begin{center}{\Large \bf EXACT SOLUTIONS OF EINSTEIN'S FIELD EQUATIONS } \\
\vskip.25in
{P. S. Negi}

{\it
Department of Physics, Kumaun University,
Nainital 263 002, India} %
\end{center}


\begin{abstract}                
 We examine various well known exact  solutions  available  in  
the literature to investigate  the  recent  criterion obtained in ref. [20]  which  
should be fulfilled by  any static and spherically symmetric  solution  in 
the  state  of   hydrostatic  equilibrium.  It  is  seen  that  this 
criterion  is   fulfilled only by (i) the regular solutions having a vanishing
surface  density together  with  the  pressure, and (ii) the singular
solutions corresponding to a non-vanishing density at the surface of  the  
configuration . On  the  other  hand, the regular solutions corresponding to
a non-vanishing surface density do not fulfill this criterion. Based  upon  
this   investigation, we point out that the exterior Schwarzschild solution
itself  provides  necessary   conditions for the types  of  the density
distributions  to  be   considered inside the mass, in order to obtain exact 
solutions or equations of state compatible  with  the   structure  of  general
 relativity. The regular   solutions with finite centre and non-zero surface
densities  which do not fulfill the criterion  [20],  in fact, can not  meet
the requirement of the `actual mass' set up by exterior Schwarzschild solution.
The only regular solution  which could be possible in this regard is
represented by uniform (homogeneous) density distribution.  
     The criterion [20] provides a necessary and sufficient  condition  for
any static and spherical configuration (including core-envelope models) to be
compatible with the structure of general relativity. Thus, it may find
application to construct the appropriate core-envelope models of stellar
objects like  neutron stars and may be used to test various   equations of
state for dense nuclear matter and  the  models  of   relativistic stellar
structures like star clusters.


\end{abstract}
\newpage

\section{ INTRODUCTION}

 The first two  exact   solution   of   Einstein's   field    equations  
were obtained by Schwarzschild [1], soon after Einstein introduced General Relativity (GR).
The first solution describes the geometry of the space-time exterior to a prefect
fluid sphere in hydrostatic equilibrium. While the other, known as interior Schwarzschild solution, corresponds to the interior geometry
of a fluid sphere of constant (homogeneous) energy-density, $E$. The importance of these two
solutions in GR is well known. The exterior solution at a given point depends only upon the 
total mass of the gravitating body and the radial distance as measured from the centre of the spherical
symmetry, and not upon the `type' of the density distribution considered
inside the mass. However, we will focus  on this point of crucial importance
later on in the present paper. On the other hand, the interior Schwarzschild
solution  provides  two  very  important  features  towards   obtaining
configurations in  hydrostatic  equilibrium,  compatible   with GR, namely -
(i) It gives an  absolute  upper   limit on compaction parameter, $u (\equiv
M/a$, mass  to  size  ratio  of   the entire configuration in geometrized
units) $\leq (4/9)$ for any static and spherical solution (provided the
density decreases monotonically outwards from the centre) in hydrostatic
equilibrium   [2], and (ii) For an assigned value of the  compaction 
parameter,   $u$,  the  minimum  central  pressure, $ P_0$,   corresponds   to
  the   homogeneous density solution (see,  e.g.,  [3]).  Regarding  these  
conditions, it should be noted that the  condition  (i)  tells  us   that the
values higher than the  limiting  (maximum)  value  of  $u   (=4/9)$ can not
be attained by any static solution. But,  what   kinds of density
variations are  possible for a mass to  be  in  the   state of hydrostatic equilibrium?,
the answer  to  this  important   question could be provided by an appropriate
analysis of the condition   (ii), and the necessary conditions put forward by
exterior Schwarzschild solution. 

      Despite the non linear differential equations, various  exact  
solutions  for  static  and  spherically  symmetric   metric   are  
available  in  the  literature  [4].  Tolman  [5]  obtained   five  
different types of exact solutions for static cases, namely - type  
III (which corresponds to the constant density  solution  obtained  
earlier by Schwarzschild [1]), type IV, type V, type VI, and  type  
VII. The solution independently obtained by Adler [6],  Adams  and  
Cohen  [7],  and  Kuchowicz  [8].  Buchdahl's  solution  [9]   for  
vanishing surface density  (the  ``gaseous''  model).  The  solution  
obtained by Vaidya  and  Tikekar  [10],  which  is  also  obtained  
independently by Durgapal and Bannerji [11]. The  class  of  exact  
solutions discussed  by  Durgapal  [12],  and  also  Durgapal  and  
Fuloria [13] solution.  Knutsen  [14]  examined  various  physical  
properties of the solutions mentioned in references ([6 - 8],  [10  
- 11], and [13]) in great detail, and found that  these  solutions  
correspond to nice physical  properties  and  also  remain  stable  
against small radial pulsations upto certain values of $u$. Tolman's  
V and VI solutions are not considered physically viable,  as  
they correspond to singular solutions [infinite values of central
density (that is, the metric coefficient, $e^{\lambda} \neq 1$ at $r =
0$) and pressure for all permissible values of $u$]. Except Tolman's V and VI
solutions,  all other solutions mentioned above are known as  regular
solutions [finite positive density at the origin (that is, the metric
coefficient, $e^{\lambda} = 1$ at $r = 0$) which decreases monotonically
outwards], which can be further divided into two categories: (i) regular solutions 
corresponding to a vanishing density at the surface together with pressure (like, 
Tolman's  VII solution (Mehra [15], Durgapal and Rawat [16],  and  
Negi   and Durgapal [17,
18]), and  Buchdahl's  ``gaseous'' solution  [9]), and (ii) regular solutions correspond
to a non-vanishing density at the surface (like, Tolman's III and IV solutions [5], and
the solutions discussed in the ref.[6 - 8], and [10 - 13] respectively).

     The stability analysis of Tolman's  VII  solution  with  
vanishing surface density has been undertaken in  detail  by  Negi  
and Durgapal [17, 18] and they have shown that this solution  also  
corresponds to  stable  Ultra-Compact  Objects  (UCOs)  which  are  
entities of physical  interest.  This  solution  also  shows  nice  
physical properties, such  as,  pressure  and  energy-density  are  
positive and finite everywhere,  their  respective  gradients  are  
negative, the ratio of pressure to density  and  their  respective  
gradients decrease outwards etc. The other solution  which  falls  
in this  category  and  shows  nice  physical  properties  is  the  
Buchdahl's  solution  [9], however, Knutsen [19]  has  shown  that  
this solution  turned  out  to  be  unstable  under  small  radial  
pulsations. 

     All these  solutions  (with  finite,  as  well  as  vanishing  
surface density) discussed above, in fact, fulfill  the  criterion  
(i), that is, the equilibrium configurations pertaining  to  these  
solutions always correspond to a value of compaction parameter, $u$,  
which is always less than the Schwarzschild  limit,  i.  e.,  $u \leq  
(4/9)$, but, this condition alone  does  not  provide  a  necessary  
condition for hydrostatic equilibrium. Nobody has discussed until  
now, whether these solutions also fulfill  the  condition  (ii)  ?  
which is necessary to satisfy by any static and spherical configuration in
the state of hydrostatic equilibrium.   

     Recently, by using the condition (ii),  we  have connected  the compaction
parameter, $u$, of any static and spherical configuration with the
corresponding ratio of  central pressure  to central energy-density  $\sigma
[\equiv (P_0/E_0)$] and worked out  an   important criterion which concludes
that for a given value of $\sigma$, the maximum   value   of compaction
parameter, $u(\equiv u_h)$, should always correspond to  the  homogeneous  
density sphere [20]. An examination of this criterion on some well   known
exact solutions and equations of state (EOSs) indicated  that   this criterion,
in fact, is fulfilled only by those configurations   which correspond to a
vanishing density at  the  surface  together   with pressure [20], or by
the singular solutions with non-vanishing surface density [section 5 of the
present study]. This result has motivated us  to  investigate,   in  detail, 
the  various  exact  solutions   available   in   the   literature, and
disclose the reason (s) behind non-fulfillment  of   the said criterion by
various regular analytic solutions and EOSs   corresponding to a
non-vanishing finite density at the surface  of   the configuration. In this
connection, in the present paper, we have examined  various   exact solutions
available in the literature in detail. It is  seen   that Tolman's VII
solution  with  vanishing  surface  density   [15, 17, 18], Buchdahl's
``gaseous'' solution [9], and Tolman's V and VI singular solutions pertain  to 
a   value of $u$ which always turns out to be less than the value, $u_h$, of
the homogeneous density sphere for all assigned values of $\sigma$. On   the
other hand, the solutions  having  a  finite  non-zero   surface density (that
is, the  pressure  vanishes  at  the  finite   surface density) do not show
consistency with the structure of the   general relativity, as they correspond
to a value of $u$ which turns   out to be greater than $u_h$  for all assigned
values of $\sigma$,  and  thus   violate the criterion obtained in [20]. 

     One may ask,  for  example,  what  could  be,  in  fact,  the  
reason(s) behind non-fulfillment of the criterion obtained in [20]  
by various exact solutions  (corresponding  to a finite,  non-zero  
density at the surface) ? We  have  been  
able  to  pin  point (which is discussed under section 3 of the present study)
 the  main  reason,  namely, the `actual' total mass $`M'$ which appears in
the exterior Schwarzschild solution, in fact,  can not be attained by the
configurations corresponding to a regular density variation with non-vanishing
surface density.

\section{FIELD EQUATIONS AND EXACT SOLUTIONS}

The metric inside a static and spherically  symmetric  mass  distribution 
corresponds to 
\begin{equation} ds^2  = e^\nu dt^2  - e^\lambda dr^2  - r^2 d\theta^2 - r^2\sin^2\theta d\phi^2, \end{equation}
where $\nu$ and $\lambda$ are functions of $r$ alone. The resulting field  equations
for the metric governed by Eq.(1) yield in the following form
\begin{eqnarray} 8\pi T^0_{0} & = & 8\pi E = e^{-\lambda} [(\lambda'/r) - (1/r^2 )] + 1/r^2, \\
- 8\pi T^1_{1} & = & 8\pi P = e^{-\lambda} [(\nu'/r) + (1/r^2)] - 1/r^2, \\
- 8\pi T^2_{2} & = & - 8\pi T^3_{3} = 8\pi P = e^{-\lambda} [(\nu''/2) + (\nu'^2/4)-(\nu'\lambda'/4)+(\nu'- \lambda')/2r]. \end{eqnarray}
where the primes represent differentiation with respect to $r$, the speed of light $c$
and the Universal gravitation constant $G$ are chosen as unity (that is, we are
using the geometrized units).  $P$ 
and $E$ represent, respectively,  the  pressure  and  energy-density 
inside the perfect fluid sphere  related  with  the  non-vanishing 
components of the energy-momentum tensor, $T_i^j$ =  0,  1,  2, 
and 3 respectively.

\medskip

     Eqs.(2) - (4) represent second-order,  coupled  differential 
equations  which  can  be  reduced  to  the  first-order,  coupled 
differential equations, by eliminating $\nu''$ from Eq.  (3)  with  the 
help of Eqs. (2) and (4) in  the  well  known  form  (namely,  TOV 
equations (Tolman [5]; Oppenheimer \&  Volkoff [21]), 
governing hydrostatic equilibrium in general relativity)
\begin{equation}
P'  =  -(P + E)[4\pi Pr^3  + m(r)]/r(r - 2m(r)),                  
\end{equation}
\begin{equation}
\nu'  =  - 2P'/(P + E),                                          
\end{equation}
and
\begin{equation}
m'(r)  =  4\pi Er^2 ,                                               
\end{equation}
where prime denotes differentiation with respect to $r$, and $m(r)$
is defined as the mass-energy contained within the radius $`r'$, that is

\begin{equation}
m(r)  =  \int_{0}^{r} 4\pi Er^2 dr.                                            
\end{equation}  
The equation connecting metric parameter $\lambda$ with $m(r)$ is given by

\begin{equation}
e^{-\lambda}    =  1  -  [2m(r)/r] = 1 - (8\pi/r)\int_{0}^{r} Er^2 dr.                                   
\end{equation}

The three field equations (or TOV equations) mentioned 
above, involve four variables, namely, $P, E, \nu,$ and  $\lambda$.  Thus,  in 
order to obtain a solution of these equations, one more equation is needed [which may 
be assumed as a relation between $P$, and $E$ (EOS), or can be  regarded  as 
an algebraic relation connecting one of the  four  variables  with 
the radial coordinate $r$ (or an algebraic relation between the parameters)].
For obtaining an exact solution, the later approach is employed.  

\medskip

Notice  that Eq.(9) yields the metric coefficient $e^{\lambda}$ for the assumed
energy-density, $E$, as a function of radial distance $`r'$. Once the metric coefficient
$e^{\lambda}$ or mass $m(r)$ is defined for assumed  energy-density by using Eqs.(9) or (8), 
the pressure, $P$, and the metric coefficient, $e^{\nu}$,
can be obtained by solving Eqs.(5) and (6) respectively which yield
two constants of integration. These constants should be obtained from the following boundary 
conditions, in order to have a proper solution of the field equations:

\section{BOUNDARY CONDITIONS: THE VALID AND INVALID ASSUMPTIONS FOR MASS DISTRIBUTION 
}

(B1) In order to maintain hydrostatic  equilibrium  throughout  the 
configuration, the pressure must vanish  at  the  surface  of  the 
configuration, that is
\begin{equation}
P  =  P(r = a)  = P_a =  0,                                     
\end{equation}
where $`a'$ is the radius of the configuration.

\medskip

(B2) The consequence of Eq.(10) ensures  the  continuity  of  the  
metric parameter $e^\nu$, belonging to the interior solution with  the  
corresponding expression for  well  known  exterior  Schwarzschild  
solution at the surface  of  the  fluid  configuration,  that  is:  
$e^{\nu(r = a)} = 1 - (2M/a)$ [where $`M' = m(a)$ is the total mass  of  the  
configuration].  However,  the  exterior  Schwarzschild   solution  
guarantees that: $e^{\nu(r = a)} = e^{-\lambda(r = a)}$, which means that the  matching  
of the metric parameter $e^\lambda$  is also ensured at the surface  of  the  
configuration together with $e^\nu$, that is 
 
\begin{equation}
e^{\nu(a)}  =  e^{-\lambda(a)}  =  1  - (2M/a)  =  (1 - 2u), 
\end{equation}

irrespective of the condition that the surface density, $E_a = E(r = a)$,
is vanishing with pressure or not, that is
\begin{equation}
 E_a = 0,                                     
\end{equation}
together with Eq. (10), or
\begin{equation}
 E_a \neq 0,                                     
\end{equation}

where $`u'$
is  the  `compaction 
parameter' of the configuration defined earlier, and $`M'$ is defined as [Eq. (8)] 
 
\begin{equation}
M = m(a)  =  \int_{0}^{a} 4\pi Er^2 dr.                                            
\end{equation}        
        
Thus, the analytic solution for the fluid sphere can  be  explored  
in terms of the only free parameter $`u'$ by normalizing the  metric  
coefficient $e^\lambda$  [yielding  from  Eq.(11)]  at  the  surface  of  the  
configuration [that is, $e^\lambda = (1 - 2u)^{-1}$ at $r = a$]  after  obtaining  
the integration constants by using Eqs.(10) [that is, $P = 0$ at $r =  
a$] and (11) [$e^\nu  = (1 - 2u)$ at $r = a$] respectively. 

\medskip

However, at this place we recall the well known property of the exterior
Schwarzschild solution (which follows directly from the definition of the
mass, $M$, appears in this solution), namely - at a given point outside the
spherical distribution of mass, $M$, it depends only upon $M$, and not upon
the `type' of the density variation considered inside the radius, $a$, of this
sphere. It follows, therefore, that the dependence of mass, $M$, upon the
`type' of the density distribution plays an important role in order to fulfill
the requirement set up by exterior Schwarzschild solution. The relation, $M =
au$, immediately tells us that for an assigned value of the compaction
parameter, $u$, the mass, $M$, depends only upon the radius, $a$, of the
configuration which may either depend upon the surface density, or upon the
central density, or upon both of them, depending upon the `type' of the
density variation considered inside the mass generating sphere. We argue that
this dependence should occur in such a manner that the definition of mass,
$M$, is not violated. We infer this definition as the `type independence'
property of the mass, $M$, which may be defined in this manner: ``The mass,
$M$, which appears in the exterior Schwarzschild solution, should either
depend upon the surface density, or upon the central density, and in any case,
not upon both of them so that from an exterior observer's point of view, the
`type' of the density variation assigned for the mass should remain
unidentified''.

\medskip

We may explain the `type independence' property of mass, $M$, mentioned above in
the following manner: The mass, $M$, is called the coordinate mass, that is, the
mass as measured by some external observer, and from this observer's point of view,
if we are `measuring' a sphere of mass, $M$, we can not know, by any means, the way
in which the matter is distributed from the centre to the surface of this sphere,
that is, if we are measuring, $M$, with the help of non-vanishing surface density
[obviously, by calculating the (coordinate) radius, $a$, from the expression connecting
the surface density and the compaction parameter, and by using the relation, $M = au$],
we can not measure it, by any means, from the knowledge of the central density (because, if we
can not know, by any means, the way in which the matter is distributed from the centre to the
surface of the configuration, then how can we know about the central density?), and this is possible
only when there exist no relation connecting the mass, $M$, and the central density, that is, the
mass, $M$, should be independent of the central density, meaning thereby that {\em the surface density
should be independent of the central density for configurations corresponding to a non-vanishing
surface density}. However, if we are measuring the mass, $M$, by using the expression for central
density (in the similar manner as in the previous case, by calculating the radius, $a$, from the
expression of central density, and using the relation, $M = au$), we can not calculate it, by any
means, from the knowledge of the surface density (in view of the `type independence' property of
the mass, $M$), and this is possible only when there exist no relation connecting the mass, $M$, and
the surface density, meaning thereby that the central density should be
independent of the surface density.

\medskip

From the above explanation of `type independence' property of mass $`M'$, it
is evident that {\em the `actual' total mass $`M'$ which appears in the
exterior Schwarzschild solution should either depend upon the surface density,
or depend upon the central density of the configuration, and in any case, not
upon both of them}. However, the dependence of mass $`M'$ upon both of the
densities (surface, as well as central) is a common feature observed among all
regular solutions having a non-vanishing density at the surface of the
configuration [see, for example, Eqs.(21), (25), (29), and (33) respectively,
belonging to the solutions of this category which are discussed under
sub-sections (a) - (d) of section 5 of the present study]. Thus, it is evident
that the surface density of such solutions is {\em dependent} upon the central
density and vice-versa, that is, the total mass, $M$, depends upon {\em both}
of the densities, meaning thereby that {\em the `type' of the density
distribution considered inside the sphere of mass, $M$, is known to an
external observer} which is the violation of the definition of mass, $M$
(defined as the `type independence' property of mass $`M'$ above), such
structures, therefore, do not correspond to the `actual' total mass $`M'$
required by the exterior Schwarzschild solution to ensure the condition of
hydrostatic equilibrium. This also explains the reason behind non-fulfillment
of the `compatibility criterion' by them which is discussed under section 5 of
the present study. However, it is interesting to note here that there could
exist only one solution in this regard for which the mass $`M'$ depends upon
both, but the same, value of surface and centre density, and for regular
density distribution the structure would be governed by the homogeneous
(constant) density throughout the configuration (that is, the homogeneous
density solution).

\medskip

Note that the requirement,
`type independence' of the mass would be obviously fulfilled by the regular structures 
corresponding to a vanishing density at the surface together with pressure,  
because the mass $`M'$ will
depend only upon the central density (surface density is always zero for
these structures) [see, for example, Eqs.(37) and (41), discussed under sub-section
(e) and (f) for Buchdahl's "gaseous" model and Tolman's VII solution
having a vanishing density at the surface, respectively].

\medskip

Furthermore, the demand of `type independence' of mass, $M$, is also satisfied
by the singular solutions having a non-vanishing density at the surface,
because such structures correspond to an infinite value of central density, and
consequently,  the mass $`M'$ will depend only upon the surface density [see, for
example, Eqs.(46) and (50), discussed under sub-section (g) and (h) for Tolman's
V and VI solutions, respectively]. Both types of these structures are also
found to be consistent with the `compatibility criterion' as discussed under
section 5 of the present study.

\medskip
 
The discussion regarding various types of density distributions considered
above is true for any single analytic solution or equation of state comprises
the whole configuration. At this place, we are not intended to claim that the
construction of a regular structure with non-vanishing surface density is
impossible. It is quite possible, provided we consider a two-density
structure in such a manner that the mass $`M'$ of the configuration
turns out to be independent of the central density so that the property `type
independence' of the mass $`M'$ is satisfied. Examples of such two-density
models are also available in the literature (see, e.g., ref. [22]), but in the
different context. However, it should be noted here that {\em the fulfillment
of `type independence' condition by the mass $`M'$ for any two-density model
will represent only a necessary condition for hydrostatic equilibrium, unless
the `compatibility criterion' [20] is satisfied by them, which also assure a
sufficient and necessary condition for any structure in hydrostatic
equilibrium} (this issue is addressed in the next section of the present
study).

\medskip

The above discussion can be summarized in other words as: {\em although  
the  exterior  Schwarzschild  
solution itself does not depend  upon  the  type  of  the  density  
distribution or  EOS  considered  inside  a fluid sphere in the state of hydrostatic
 equilibrium,  
however, it puts the important condition that only two types of the density  
variations are possible inside the configuration in  order  to  
fulfill the condition of hydrostatic equilibrium: (1) the surface  
density of the configuration should be independent of the central density, and
(2) the  central density of the
configuration should be independent of the surface density}. Obviously, the
condition (1) will be satisfied by the configurations pertaining to an
infinite value of the central density (that is, the singular solutions),
and/or by the two-density (or multiple-density) distributions corresponding
to a surface density which turns out to be independent of the central
density (because, the regular configurations governed by a single exact
solution or EOS pertaining to this category are not possible). Whereas, the
condition (2) will be fulfilled by the configurations corresponding to a
surface density which vanishes together with pressure [the configurations in
this category will include: the density variation governed by a single exact
solution or EOS, as well as the two-density (or multiple-density)
distributions]. However, the point to be emphasized here is that a
two-density distribution in any of the two categories mentioned here will
fulfill only a {\em necessary} condition for hydrostatic equilibrium unless
the `compatibility criterion' [20] is satisfied by them which also assure a
necessary and sufficient condition for any structure in the state of
hydrostatic equilibrium as mentioned above.

 \section{CRITERION FOR STATIC SPHERICAL CONFIGURATIONS TO BE CONSISTENT
          WITH THE STRUCTURE OF GENERAL RELATIVITY}

The criterion obtained in [20] can be summarized in the  following  
manner: 
For an assigned value of the ratio of central pressure to  central  
energy-density $ \sigma[\equiv(P_0 /E_0)]$,   the   compaction   parameter    of  
homogeneous density distribution, $u(\equiv u_h)$ should always  be  larger  
than or equal to the compaction parameter $u(\equiv u_v)$  of  any  static and spherical  
solution, compatible with the  structure  of  General  Relativity.  
That is 
\begin{equation}
u_h \geq u_v  ({\rm \,for \,an \,assigned \,value \,of\,} \sigma).         
\end{equation}
In the light of Eq. (15), let us assign the same value, $M$, for  the  
total mass corresponding  to  various static  configurations  in  
hydrostatic  equilibrium.  If  we  denote  the  density   of   the  
homogeneous sphere by $E_h$ , we can write 
\begin{equation}
E_h   =  3M/(4\pi {a_h}^3)                       
\end{equation}
where $a_h$  denotes the radius of the homogeneous density sphere. 
If $a_v$  represents the radius of any other regular  sphere  for  the  
same mass, $M$, the average density, $E_v$,  of  this  configuration  would  
correspond to 
\begin{equation}
E_v   =  3M/(4\pi {a_v}^3).                                              
\end{equation}
Eq. (15) indicates that $a_v \geq a_h$. By the use of Eqs. (16) and  (17)  
we find that 
\begin{equation}
E_v  \leq E_h.                                                     
 \end{equation}
That is, for an assign value of $\sigma$, the  average  energy-density  of  
any static spherical configuration, $E_v$, should always be less than or
equal   to the density, $E_h$ , of the homogeneous density sphere for the same
  mass, $M$. 

\medskip

     Although,   the   regular    configurations    with    finite  
non-vanishing surface densities, represented by a  single  density  
variation  can  not  exist, because for such configurations the necessary condition set up by
exterior Schwarzschild solution can not be satisfied. however, we can construct
regular configurations composed of core-envelope   models  corresponding to a finite   central   
with vanishing and  non-vanishing surface densities, such that the necessary conditions imposed by 
the Schwarzschild's exterior solution
at the surface of the configuration are appropriately satisfied. However, it should be 
noted that the necessary conditions satisfied by such core-envelope models at the surface
may not always turn out to be sufficient for describing the state of hydrostatic equilibrium
[because for an assigned value of $\sigma$, the average density of such configurations may not always turn out to be 
less than or equal to the density of the homogeneous density sphere for the same mass, as indicated
by Eqs.(16) and (17) respectively (it would depend upon the types of the
density variations considered for the core and envelope regions and the the
matching conditions at the core-envelope boundary)]. Thus, it follows that the
criterion obtained in [20] is able to provide a {\em necessary} and {\em
sufficient} condition for any regular configuration to be consistent with the
state of hydrostatic equilibrium.

\medskip

The future study of  such core-envelope models [see, for example, the models
described in [22] and [23]],  based upon the  criterion  obtained  in   [20] 
could  be  interesting  regarding  two density structures of neutron stars and
other stellar objects compatible with the structure of GR.   
 \section{EXAMINATION OF THE COMPATIBILITY CRITERION FOR VARIOUS WELL KNOWN
           EXACT SOLUTIONS AVAILABLE IN THE LITERATURE}  

     We have considered the following exact solutions expressed in  
units of compaction parameter $u [\equiv (M/a)$, mass  to  size  ratio  in  
geometrized units], and radial coordinate  measured  in  units  of  
configuration  size,  $y  [\equiv (r/a)]$,  for  convenience.  The   other  
parameters which will appear in these solutions,  are  defined  at  
the relevant places.  In  these  equations,  $P$  and  $E$  represent,  
respectively,  the  pressure   and   energy-density   inside   the  
configuration. The surface density is denoted by $E_a$, and the central
pressure and central energy-density are   denoted by $P_0$, and $E_0$,
respectively.   

\medskip

     The regular exact solutions which pertain  to a non-vanishing  value  of  
the surface density are given under the sub-sections (a) -  (d),  while  
those correspond to a vanishing  value  of  the  surface  
density are described under the sub-sections (e) and (f) respectively.
Sub-sections (g) and (h) represent the singular solutions having
non-vanishing values of the surface densities.   
\medskip

(a) Tolman's IV solution
\begin{equation}
8\pi Pa^2 = \frac{3u^2(1 - y^2)}{(1 - 3u + 2uy^2)},
\end{equation}
\begin{equation}
8\pi Ea^2 = \frac{u(4 - 9u + 3uy^2)}{(1 - 3u + 2uy^2)} + \frac{2u(1 - 3u)(1 - uy^2)}{(1 - 3u + 2uy^2)^2}.
\end{equation} 
By the use of Eq.(20), we can obtain the relation connecting central and
surface densities in the following form
\begin{equation}
(E_a/E_0) = \frac{2(1 - 2u)(1 - 3u)}{(1 - u)(2 - 3u)}.
\end{equation}
Eq.(21) shows that the surface density is {\em dependent} upon the
central density, and vice-versa.

\medskip

  By using Eqs. (19) and (20), we obtain 
\begin{equation} 
(P_0/E_0)  =  u/(2 - 3u). 
\end{equation}                           
 
This solution finds application [it  can  be  seen  from  Eq.  
(19)] for the values of $u \leq (1/3)$. 
 
\bigskip
 
(b) Adler [6], Adams and Cohen [7], and Kuchowicz [8] solution: 
\begin{equation}
8\pi Pa^2 = \frac{2}{(1 + z)} \Bigl [2x - \frac{u(1 + 5z)(1 + 3x)^{2/3}}{(1 + 3z)^{2/3}} \Bigr ],
\end{equation}
\begin{equation}
8\pi Ea^2 = \frac{2u(3 + 5z)(1 + 3x)^{2/3}}{(1 + 5z)^{5/3}},
\end{equation}                                          
 
where $z = xy^2$, and $u = 2x/(1 + 5x)$.

Eq.(24) gives the relation, connecting central and surface densities of the
configuration in the following form
\begin{equation}
(E_a/E_0) = \frac{(3 + 5x)}{3(1 + 5x)^{5/3}}.
\end{equation}
Thus, the surface density {\em depends} upon the central density
 and vice-versa.

  \medskip

Equations (23) and (24) give 
  
\begin{equation} 
(P_0 /E_0)  =  (1/3) \Bigl [\frac{(1 + 5x)}{(1 + 3x)^{2/3}} - 1 \Bigr ].
\end{equation}
  
It is seen from Eq.(23) that this solution finds application  
for values of $u \leq (2/5)$. 

\bigskip

(c) Vaidya and  Tikekar  [10],  and  Durgapal  and  Bannerji  [11]  
solution 
\begin{equation}                
8\pi P  =  (9C/2) \frac{(1 + x)^{1/2}(1 - x) - B(1 + 2x)(2 - x)^{1/2}}{(1 + x)[(1 + x)^{3/2} + B(2 - x)^{1/2}(5 + 2x)]},
\end{equation}
 
\begin{equation} 
8\pi E  =  (3C/2) \frac{(3 + x)}{(1 + x)^2},
\end{equation}
 
where the variable $x$, and the constants $B$ and $C$ are given by                      

\medskip
                                               
$x = Cr^2; X = Ca^2 = 4u/(3 - 4u)$; and 

\[B = \frac{(1 + X)^{1/2}(1 - X)}{(1 + 2X)(2 - X)^{1/2}}.\]
By the use of Eq.(28), we find that the surface and central densities are
connected by the following relation
\begin{equation}
(E_a/E_0) = \frac{(3 + X)}{3(1 + X)^2}.
\end{equation}
It is evident from Eq.(29) that the surface density is {\em dependent} upon the
central density, and vice-versa.

\medskip

   By the use of Eqs.(27) and (28), we obtain 

\begin{equation} 
(P_0/E_0)  =  \frac{(1 - B\sqrt{2})}{(1 + 5B\sqrt{2})}.
\end{equation}
 
This solution finds application for the values of $u \leq 0.4214$,  as  
shown by Eq.(27). 

\bigskip

(d) Durgapal and Fuloria [13] solution 
\begin{equation} 
\frac {8\pi P}{C}  =  \frac{16}{7z(1 + x)^2} \Bigl [(1 + x)(2 - 7x - x^2) - A[(18 + 25x + 3x^2)(7 - 10x - x^2)^{1/2} - 4(1 + x)(2 - 7x - x^2) W(x)] \Bigr ],
\end{equation}
\begin{equation}                                                                
\frac {8\pi E}{C}  =  \frac {8(9 + 2x + x^2)}{7(1 + x)^3}.
\end{equation}
 
where $C$ is a constant and $x = Cr^2$. 

\medskip
The variables $z$ and  $W(x)$ are given by 

\[z  =  (1 + x)^2  + A[(7 + 3x)(7 - 10x - x^2)^{1/2} + 4W(x)(1 + x)^2], \] 
                                   
\[W(x)  =  {\rm ln} \Bigl [ \frac{(3 - x) + (7 - 10x - x^2)^{1/2}}{(1 + x)} \Bigr ], \]

where the arbitrary constant $A$ is given by 
                                           
\[A = \frac{(1 + X)(2 - 7X - X^2)}{[(18 + 25X + 3X^2)(7 - 10X - X^2)^{1/2} - 4(1 + X)(2 - 7X - X^2)W(X)]} \]
 
and $X  =  Ca^2$.

Eq.(32) gives the relation, connecting central and surface densities as
\begin{equation}
(E_a/E_0) = \frac{(9 + 2X + X^2)}{9(1 + X)^3}.
\end{equation}
Eq.(33) indicates that the surface density is {\em dependent} upon the central
density, and vice-versa.

\medskip 
By the use of equations (31) and (32), we get 
 
\begin{equation}
(P_0/E_0) = \frac {2}{9z(0)} \Bigl [2 - A[18\sqrt{7} - 8W(0)] \Bigr ],
\end{equation} 
                                          
where $z(0)$  and $W(0)$ are given by 
 
\[z(0)  =  1  +  A[7\sqrt{7}  +  4W(0)], \]
 
\[W(0)  =  {\rm ln}(3 + \sqrt{7}),\] 
and 
 
\[u  =  \frac{8X(3 + X)}{14(1 + X)^2}.\]
                     
As seen from Eq.(31), this solution is applicable  for  the  
values of $u \leq 0.4265$.

\bigskip
(e) Buchdahl's ``gaseous'' solution [9] 
\begin{equation}
8\pi Pa^2 = \frac{\pi^2(1 - u)^2}{(1 - 2u)} \frac{m^2}{(n + m)^2},
\end{equation}
 
\begin{equation}             
8\pi Ea^2 = \frac{\pi^2(1 - u)^2}{(1 - 2u)} \frac{m(2n - 3m)}{(n + m)^2},
\end{equation}
 
where $m  = 2u[sin(z)/z]; \,\, \, n  =  2(1 - u)$,

\medskip
and

\[z = \frac {\pi y}{[1 + u(sin(z)/z)(1 - u)^{-1}]};\,\,\, 0 \leq z \leq \pi\]

Eq.(36) shows that the surface density vanishes together with pressure,
thus, the central density will become {\em independent} of the surface density,
given by the equation
\begin{equation}
8\pi E_0a^2 = \frac{{\pi}^2u(2 - 5u)(1 - u)^2}{(1 - 2u)}.
\end{equation}

 By using equations (35) and (36), we obtain  \begin{equation} 
(P_0 /E_0 )  =  u/(2 - 5u).                                     
\end{equation}
It is evident from Eq.(35) that  this  solution  is  
applicable for the values of $u \leq (2/5)$. 

\bigskip

(f) Tolman's VII solution with vanishing surface density 
\begin{equation} 
8\pi Pa^2  =  La^2 \frac {C_2\, {\rm cos}(w/2) - C_1\, {\rm sin}(w/2)}{C_1\, {\rm cos}(w/2) + C_2\, {\rm sin}(w/2)} - Na^2,
\end{equation}
\begin{equation}
8\pi Ea^2  =  8\pi E_0a^2(1 - x),                 
\end{equation}
where $E_0$  is the central energy-density given by    
\begin{equation}
8\pi E_0a^2  = 15u.
\end{equation}
and 
\[x = (r/a)^2 = y^2,\]  
\[C_1 = A\, {\rm cos}(w_a/2) - B\, {\rm sin}(w_a/2);\,\, C_2 = A\, {\rm sin}(w_a/2) + B\, {\rm cos}(w_a/2), \]
\[La^2 = 2(3u)^{1/2}[1 - ux(5 - 3x)]^{1/2};\,\, Na^2 = u(5 - 3x),\]
\[w = {\rm ln}[x - (5/6) + ([1 - ux(5 - 3x)]/3u)^{1/2}], \]
\[w_a = ({\rm the\,\,value\,\,of\,\,w\,\,at}\,\,y = 1) = {\rm ln}[(1/6) + [(1 - 2u)/3u]^{1/2}],\]
\[A = (1 - 2u)^{1/2};\,\, B = (u/3)^{1/2}.\]

By using Eqs. (39) and (40), we get 
\begin{equation} 
(P_0 /E_0) = (1/3) \Bigl [(2/5)(3/u)^{1/2} \frac {C_2\, {\rm cos}(w/2) - C_1\, {\rm sin}(w/2)}{C_1\, {\rm cos}(w/2) + C_2\, {\rm sin}(w/2)} -  1 \Bigr ],
\end{equation}                                                              
where $w_0$  is given by 
\[w_0   =  {\rm ln}[(1/3u)^{1/2} - (5/6)].\]

It follows from Eq.(40) that the surface density is always zero, hence the
central density is always {\em independent} of the surface density.

\medskip

Eq.(39) indicates that his solution is applicable for the values of $u  \leq 
0.3861$. 

\bigskip 
(g) Tolman's V solution
\begin{equation}
8\pi Pa^2 = \frac {n^2(1 - y^q)}{(2n + 1 - n^2)y^2},
\end{equation}
and

\begin{equation}
8\pi Ea^2 = \frac {n\Bigl [(2 - n) + [n(3 - n)/(1 + n)]y^q \Bigr ]}{(2n + 1 - n^2)y^2},
\end{equation}
where $q$ is given by

\[q = 2(2n + 1 - n^2)/(1 + n), \]
and $n$ is defined as

\[n = u/(1 - 2u). \]

Eq.(44) shows that the central density is always infinite (for $n < 2$) together with
central pressure (Eq.43), however, their ratio ($P/E$) is finite at all points inside the
configuration, and at the centre, yields in the following form 
\begin{equation}
(P_0/E_0) = n/(2 - n) = u/(2 - 5u).
\end{equation}
The consequence of the infinite central density is that the surface
density will become {\em independent} of the central density, given
by the equation
\begin{equation}
8\pi E_aa^2 = 2n/(n + 1) = 2u/(1 - u).
\end{equation}

It is evident from Eq.(45) that this solution is applicable for a value of $n
\leq 2$ [i. e., for a value of $u \leq (2/5)$].

\bigskip

(h) Tolman's VI solution

\begin{equation}
8\pi Pa^2 = \frac {{(1 - n^2)}^2 (1 - y^{2n})}{y^2(2 - n^2)[(1 + n)^2 -
y^{2n}(1 - n)^2]}, 
\end{equation}

\begin{equation}
8\pi Ea^2 = \frac {(1 - n^2)}{y^2(2 - n^2)}.
\end{equation}
Eqs.(47) and (48) indicate that the central pressure and central density are always infinite, 
however, their ratio ($P/E$) is finite at all points inside the structure, and at the centre, 
reduces into the following form 
\begin{equation}
(P_0/E_0) = (1 - n^2)/{(1 + n)}^2 = (1 - n)/(1 + n),
\end{equation}
and the
surface density (obviously, {\em independent} of the central density) would be
given by the equation
\begin{equation}
8\pi E_aa^2 = \frac{(1 - n^2)}{(2 - n^2)} = 2u.
\end{equation}
where $n$ is defined as

\[n^2 = (1 - 4u)/(1 - 2u). \]

Eq.(49) indicates that this solution is applicable for a value of $u
\leq (1/4)$.

\medskip
 
     Let us denote the compaction parameter of the homogeneous density  
configuration by $u_h$ , and for the exact solutions corresponding  to  
the sub-sections (a) - (d) by $u_{IV}, u_{AAK}, u_{DBN},$ and $u_{DFN}$ 
respectively.  The   compaction  parameters  of the exact   solutions  
described   under   sub-section (e) and (f) are denoted by $u_{BDL}$ and
$u_{VII}$ respectively, and those discussed under sub-sections (g) and (h) are
denoted by $u_{V}$ and $u_{VI}$ respectively.

\medskip

     Solving these analytic solutions for various assigned  values  
of the ratio of central pressure to central  energy-density $\sigma [\equiv  
(P_0 /E_0)]$, we obtain the  corresponding  values  of  the  compaction  
parameters as shown in Table 1 and Table 2 respectively. It is seen that for
each  and every assigned   value of $\sigma$, the values  represented by
$u_{IV}, u_{AAK}, u_{DBN},$ and $u_{DFN}$ respectively (Table 1),  turn  
out to be higher than $u_h$  (that is, $u_{IV}, u_{AAK}, u_{DBN},$ and $u_{DFN}
\geq u_h$), while those represented by $u_{BDL}, u_{VII}, u_{V}$, and
$u_{VI}$ respectively (Table 2) correspond to a value
which always  remains   less than $u_h$  (that is, $u_{BDL}, u_{VII},
u_{V}$, and $u_{VI} \leq  u_h$). Thus,   we conclude that the configurations
defined by $u_{IV}, u_{AAK}, u_{DBN},$ and $u_{DFN}$ respectively, do  not 
show  compatibility  with  the   structure   of   general   relativity, while
those defined by $u_{BDL}, u_{VII}, u_{V}$, and
$u_{VI}$ respectively,    show   compatibility with the
structure of general  relativity. However,   this type of characteristics
[that is,  the  value  of  compaction   parameter larger than the value of
$u_h$  for  some  or  all  assigned   values of $\sigma$] can be seen for any 
regular  exact  solution  having  a   finite non-vanishing surface density
[because such exact solutions   (having finite  central  densities)  with 
non-vanishing   surface densities can not possess the actual mass, $M$,
required to fulfill the boundary conditions at the surface]. On the other
hand, the value of compaction parameter for a regular
solution  with   vanishing surface density, and a singular solution with
non-vanishing surface density will always remain less than the value of $u_h$
for all assigned values of $\sigma$ [because such solutions naturally
fulfill the definition of the actual mass, $M$, required for the hydrostatic
equilibrium].  

\medskip
 
  Therefore, it is evident  that  the findings based upon the
`compatibility criterion' carried out in this section are fully consistent
with the definition of the mass, $M$ (defined as the `type independence'
property under section 3 of the present study).

\section{RESULTS AND CONCLUSIONS}     

 We have investigated the criterion
obtained in the  reference   [20] which states : for an assigned value of the
ratio of  central   pressure to  central  energy-density,  $\sigma (\equiv P_0
/E_0)$,  the  compaction   parameter, $u(\equiv M/a)$, of any static and
spherically symmetric solution should always be  less   than or equal to the
compaction parameter, $u_h$, of the  homogeneous   density distribution. We
conclude that this criterion is fully consistent   with the reasoning
discussed under section 3 which states that in order to fulfill the
requirement set up by exterior Schwarzschild solution (that is, to ensure
the condition of hydrostatic equilibrium), the total mass, $M$, of the
configuration should depend either upon the surface density (that is,
independent of the central density), or upon the central density (that is,
independent of the surface density), and in any case, not upon both of them.

\medskip

     An examination, based upon this criterion,  show  that  among  
various exact solutions of the field  equations  available  
in  the  literature,  the regular solutions  corresponding  to a vanishing  
surface density together with pressure, namely - (i) Tolman's  
VII solution with vanishing surface density, and  (ii)  Buchdahl's  
``gaseous'' solution, and the singular solutions with non-vanishing surface
density, namely - Tolman's V and VI solutions are compatible with the 
structure  of  general   relativity. The only regular solution  with  finite 
non-vanishing   surface density which could exist in this regard is  described
 by   constant (homogeneous) density distribution.   

\medskip
 
	This criterion provides a necessary and sufficient condition for any
static spherical configuration to be compatible with the structure of
general relativity, and may be used to construct  core-envelope  models   of
stellar  objects  like  neutron  stars  with  vanishing  and   non-vanishing
surface densities, such that for an  assigned  value   of central pressure to
central density, the average density of the   configuration should always
remain  less  than  or  equal  to  the   density of the homogeneous sphere for
the same mass.

\medskip   

     This criterion could provide a convenient and  reliable  tool  
for testing equations of state (EOSs) for dense nuclear matter and  
models of relativistic  star  clusters,  and  may find application to  
investigate new analytic solutions and EOSs.

\acknowledgments The author acknowledges  State  Observatory, 
Nainital for providing library and computer-centre facilities.

\begin{table*}     
\caption{ Various values (round off at the fourth
decimal place) of the compaction parameter $u(\equiv M/a)$ as
obtained for different assigned values of the ratio of the centre pressure to
centre energy-density, $\sigma[\equiv (P_0/E_0)]$,  corresponding to the regular exact
solutions with finite non-vanishing surface densities, namely - Tolman's
IV [5] solution [indicated by $u_{IV}$], Adlar [6], Adams and Cohen [7], and
Khchowicz's [8] solution [indicated by $u_{AAK}$], Vaidya and Tikekar [10],
and Durgapal and Bannerji [11] solution [indicated by $u_{DBN}$], and Durgapal
and Fuloria [13] solution [indicated by $u_{DFN}$] respectively.  The
compaction parameter corresponding to homogeneous density distribution
(Schwarzschild's interior solution) is indicated by $u_h$ for the same value
of $\sigma$. It is seen that for each and every assigned value of $\sigma$,
 $u_{IV}, u_{AAK}, u_{DBN}$, and $u_{DFN} > u_h$ which is the
evidence that the regular solutions corresponding to a finite non-vanishing surface density
 (indicated by $u_{IV},
u_{AAK}, u_{DBN}$, and $u_{DFN}$ respectively) are not compatible with the structure of general 
relativity.}
\begin{center}
\begin{tabular}{cccccc}


${\sigma(\equiv P_0 / E_0)}$ & $u_h$    & $u_{IV}$  & $u_{AAK}$ & $u_{DBN}$  & $u_{DFN}$ \\

0.1252  & 0.1654 & 0.1820 & 0.1745 & 0.1743 & 0.1718  \\

0.1859  & 0.2102 & 0.2387 & 0.2242 & 0.2221 & 0.2187  \\

0.2202  & 0.2301 & 0.2652 & 0.2463 & 0.2429 & 0.2392  \\

0.2800  & 0.2580 & 0.3043 & 0.2771 & 0.2714 & 0.2676   \\

0.3150  & 0.2714 & 0.3239 & 0.2917 & 0.2847 & 0.2809   \\

(1/3)   & 0.2778 & (1/3)  & 0.2984 & 0.2909 & 0.2872   \\

0.3774  & 0.2914 & & 0.3127 & 0.3038 & 0.3003  \\

0.4350  & 0.3062 & & 0.3277 & 0.3176 & 0.3145  \\

0.4889  & 0.3178 & & 0.3390 & 0.3281 & 0.3253  \\

0.5499  & 0.3289 & & 0.3493 & 0.3378 & 0.3354  \\

0.6338  & 0.3415 & & 0.3600 & 0.3485 & 0.3465  \\

0.6830  & 0.3476 & & 0.3650 & 0.3535 & 0.3519  \\

0.7044  & 0.3501 & & 0.3669 & 0.3555 & 0.3541   \\

0.7085  & 0.3506 & & 0.3673 & 0.3559 & 0.3545  \\

0.7571  & 0.3558 & & 0.3711 & 0.3601 & 0.3589  \\

0.8000  & 0.3599 & & 0.3740 & 0.3633 & 0.3624   \\

0.8360  & 0.3630 & & 0.3762 & 0.3658 & 0.3650   \\


\end{tabular}
\end{center}
\end{table*}

\begin{table*}

\caption{ Various values (round off at the fourth
decimal place) of the compaction parameter $u(\equiv M/a)$ as
obtained for different assigned values of the ratio of the centre pressure to
centre energy-density, $\sigma[\equiv (P_0/E_0)]$, corresponding to the regular 
exact solutions with vanishing
surface densities, namely - Buchdahl's [9] ``gaseous''
solution  and Tolman's VII solution [5, 15, 16, 17, 18] (indicated
by $u_{BDL}$ and $u_{VII}$ respectively), and singular solutions with non-vanishing
surface densities, namely - Tolman's V and VI solutions (indicated by $u_{V}$ and $u_{VI}$
respectively). The compaction parameter corresponding to
homogeneous density distribution (Schwarzschild's interior solution) is
indicated by $u_h$ for the same value of $\sigma$. It is seen that for each and every assigned
value of $\sigma, u_{BDL}, u_{VII}, u_{V}$, and $u_{VI} < u_h$, which is the
evidence that the regular solutions corresponding to a vanishing value of the
surface density (represented by $u_{BDL}$ and $u_{VII}$ respectively), and singular solutions
having a non-vanishing value of the surface density (represented by $u_{V}$ and $u_{VI}$ respectively) are
compatible with the structure of general relativity.}

\begin{center}
\begin{tabular}{cccccc}


${\sigma(\equiv P_0 / E_0)}$ & $u_h$    & $u_{BDL}$  & $u_{VII}$  & $u_{V}$  
&  $u_{VI}$   \\

0.1252  & 0.1654 & 0.1540 & 0.1588 & 0.1540 & 0.1417 \\

0.1859  & 0.2102 & 0.1927 & 0.1992 & 0.1927 & 0.1730 \\

0.2202  & 0.2301 & 0.2096 & 0.2166 & 0.2096 & 0.1858 \\

0.2800  & 0.2580 & 0.2333 &0.2407  & 0.2333 & 0.2031 \\

0.3150  & 0.2714 & 0.2446 &0.2521  & 0.2447 & 0.2108 \\

(1/3)   & 0.2778 & 0.2500 & 0.2574 & 0.2500 & 0.2143 \\

0.3774  & 0.2914 & 0.2614 & 0.2687 & 0.2614 & 0.2216 \\

0.4350  & 0.3062 & 0.2740 & 0.2809 & 0.2740 & 0.2290 \\

0.4889  & 0.3178 & 0.2839 & 0.2904 & 0.2839 & 0.2344 \\

0.5499  & 0.3289 & 0.2934 & 0.2993  & 0.2934 & 0.2390 \\

0.6338  & 0.3415 & 0.3040 & 0.3092  & 0.3041 & 0.2436 \\

0.6830  & 0.3476 & 0.3094 & 0.3140  & 0.3094 & 0.2455 \\

0.7044  & 0.3501 & 0.3115 & 0.3160  & 0.3115 & 0.2462 \\

0.7085  & 0.3506 & 0.3119 & 0.3164  & 0.3120 & 0.2463 \\

0.7571  & 0.3558 & 0.3164 &0.3204  & 0.3164 & 0.2476 \\

0.8000  & 0.3599 & 0.3200 & 0.3235  & 0.3200 & 0.2485 \\

0.8360  & 0.3630 & 0.3228 & 0.3260  & 0.3228 & 0.2490 \\


\end{tabular}
\end{center}
\end{table*}

\end{document}